# Proceedings to the 25th International Workshop "What Comes Beyond the Standard Models", July 4 -- July 10, 2022, Bled, Slovenia

The Members of the Organizing Committee of the International Workshop "What Comes Beyond the Standard Models", Bled, Slovenia, state that the articles published in The Proeedings to the 25th International Workshop "What Comes Beyond the Standard Models", Bled, Slovenia are refereed at the Workshop in intense in-depth discussions.

The whole Proceedings in pdf format can be found also in the Proceedings and in the Virtual [Institute of Astroparticle Physics (CosmoVia).](#)

## Contents:
• Preface in English Language
• Predgovor (Preface in Slovenian Language)
• Invited Talks
• Scientific-Educational Platform of Virtual Institute of Astroparticle Physics and
  Studies of Physics Beyond the Standard Model (by M.Yu. Khlopov)
• A poem

## Preface in English Language

This year our series of workshops on "What Comes Beyond the Standard Models?" took place the twenty-fifth time. The series started in 1998 with the idea of organizing a workshop, in which participants would spend most of the time in discussions, confronting different approaches and ideas. The picturesque town of Bled by the lake of the same name, surrounded by beautiful mountains and offering pleasant walks, was chosen to stimulate the discussions.
The idea was successful and has developed into an annual workshop, which is taking place every year since 1998. Very open-minded and fruitful discussions have become the trade-mark of our workshop, producing several published works. It takes place in the house of Plemelj, which belongs to the Society of Mathematicians, Physicists and Astronomers of Slovenia.
Since the workshop is celebrating its anniversary, 25th , it is an opportunity to look at what has happened in the field of physics of elementary fermion and boson fields all this time. And what new the measurements together with the suggested theories have brought.
The technology and computer science has progressed in mean time astonishingly, enabling almost unbelievable measurements in all fields of physics, especially in the physics of fermion and boson fields and in cosmology.

The standard model assumptions have been confirmed without offering surprises.

The last unobserved field assumed by the standard model as a field, the Higgs's scalar, detected in June 2012, was confirmed in March 2013. New and new measurements of masses of quarks and leptons and antiquarks and antileptons, of the mixing matrices of quarks and of leptons, of the Higgs's scalar, of bound states of quarks and leptons, offer new and new data. The waves of the gravitational field were detected in February 2016 and again 2017.

If we look at the collection of open questions that we set ourselves at the beginning and continuously supplemented in each workshop, it shows up that we are all the time mostly looking for an answer to the essential question: What is the next step beyond the standard model, which would offer not only the understanding of all the assumed properties for quarks and leptons and all the observed boson fields with the Higgs scalars included, but also for the observed phenomena in cosmology, like it is the understanding of the expansion rate of the universe, of the appearance of the dark matter, of black holes with their (second quantized) quantum nature included, of the necessity of the existence of the darkenergy and many others.

When trying to understand the quantum nature of fermion and boson fields we are looking for the theory which is anomaly free and possibly renormalizable so that we would be able to predict properties of second quantized fields when proposing measurements.

If it turns out that these requirements lead to the dimension of space time which is higher than $(3+1)$ then the questions arise how does it happen that all the dimensions except the $(3 + 1)$-ones are hidden. How do consequently the internal spaces of fermion and boson fields manifest in $d = (3 + 1)$? And can such a theory be free of anomalies and renormalizable?

The observations of the galactic rotation curves are one of the strongest evidence that it is not only the ordinary matter which determines the properties of the galaxies. The direct measurements of the Dama/Libra experiment with more and more accuracy demonstrates the annual dependence of the number of events. One can hardly accept that these measurements are not connected with the dark matter needed to explain the rotational curves of galaxies and behaviour of clouds of galaxies, of whatever origin the dark matter is.

The measurements of the Hubble constant state that the universe is expanding and how fast does expend. The cosmological measurements show up that there exist the black holes.

In all our annual workshops there have appeared innovative proposals for:

i. Explaining the assumptions of the standard model, mostly with new unexplained assumptions.
ii. Suggestions of what is the dark matter.
iii. Suggestions for what causes inflation.
iv. Suggestions what does determine the internal space of fermions and bosons.
v. Suggestions how to avoid anomalies and make quantum theories of fields renormalizable.
vi. How to suggest experiments which would show the next step beyond the standard model.
vii. How does "Nature make the decision" about breaking of symmetries down to the noticeable ones, if coming from some higher dimension d? viii. Why is the metric of space-time Minkowskian and how is the choice of metric connected with the evolution of our universe(s)? .... And many others.

In our proceedings there are papers (many of them later published in journals) discussing these problems and offering suggestions how to solve them.

In the last two Covid-19 years the ZOOM workshop replaced our ordinary workshops. It must be admitted that the ZOOM meetings can not replace the real meetings where all the questions are welcome even if answers need a long time to be presented.

Talks and discussions in our workshop are not at all talks in the usual way. Each talk as well as discussions lasted several hours, divided in two hours blocks, with a lot of questions, explanations, trials to agree or disagree from the audience or speakers side.

Although also on the ZOOM way of presentations several continuations of the same talk were planned and realized, yet the presence in real is much more effective.

This last, the jubilee workshop, was partly in "real" at Bled and partly by ZOOM. The topics presented and discussed in this workshop concern all the above mentioned open problems, illustrated by the question "How to understand Nature?"

We were trying to find the answers in several steps:

• How to make the next step beyond both standard models?
• How to come beyond the standard model of the so far observed quarks
and leptons and antiquarks and antileptons, appearing in families, and
interacting with the electroweak, colour and scalar fields,
• How to explain observed cosmological phenomena?
• How can we construct the anomalies-free renormalizable theory of all the so far observed
fermions and all the so far observed boson fields?
• Can this be done as well as for bound states and scattering states of fermion and boson fields?
• How do symmetries contribute to bound states?
• Can we find the way to treat all the elementary fermion and boson fields in a unique way?
• How to find the way to treat fermion and boson fields if space-time is indeed four dimensional?
• How does "Nature make the decision" about breaking of symmetries down to the noticeable ones, if coming from some higher dimension d?
• Why is the metric of space-time Minkowskian and how is the choice of metric connected with the evolution of our universe(s)?
• What does cause the inflation of the universe?
• When does the inflation appear? After the electroweak phase transition?
• What does determine the colour scale?
• What is our universe made out of besides of the (mostly) first family baryonic matter?
• How do black holes contribute to the dark matter?
• What is the role of symmetries in Nature?
• How to make experiments and how to propose the models so that the data would not be influenced too much by the proposed model?

Most of talks are "unusual" in the sense that they are trying to find out new ways of understanding and describing the observed phenomena.

The proceedings is divided into two parts. To the first part the invited talks, which appear in time and were refereed, contribute.

To the second part, called Discussion section, the contributions are presented, which started to be intensively discussed during the workshop but need more discussions so that the authors of different contributions would agree or disagree, or which seem to the authors of different contributions that they have many common points, expressed in a different way, which might lead to new ideas or new conclusions or new collaborations.

Some of discussions, started during the workshop, are not appearing in this proceedings and might continue next year and be ready for next proceedings. The organizers are grateful to all the participants for the lively presentations and discussions and the good working atmosphere although most of participants appear virtually, what was lead by Maxim Khlopov.

The reader can find all the talks and soon also the whole Proceedings on the [official website](#) of the Workshop, and on the [Cosmovia Forum.](#)

*Norma Mankoč Borštnik, Holger Bech Nielsen, Maksim Khlopov, Astri Kleppe*
Ljubljana, December 2022

# Preface in Slovenian Language

To leto je serija delavnic z naslovom „Kako preseči oba standardna modela, kozmološkega in elektrošibkega" ("What Comes Beyond the Standard Models?") stekla že petindvajsetič. Prva delavnica je stekla leta 1998 v želji, da bi udeleženci v izčrpnih diskusijah kritično soočali različne ideje in teorije. Slikovito mestece Bled, ob jezeru z enakim imenom, obkroženo s prijaznimi hribčki, nad katerimi kipijo slikovite gore, ki ponujajo prijetne sprehode in pohode, ponujajo priložnosti za diskusije.

Ideja je bila uspešna, razvila se je v vsakoletno delavnico, ki teče že ptindvajsetič. Zelo odprte, prijateljske in učinkovite diskusije so postale "blagovna znamka" naših delavnic, ideje, ki so se v diskusijah rodile, pa so pogosto botrovale objavljenim člankom. Delavnice domujejo v Plemljevi hiši na Bledu tik ob jezeru.

Hišo je Društvu matematikov, fizikov in astronomov zapustil svetovno priznani slovenski matematik Jozef Plemelj. Letošnje jubilejno leto ponuja priložnost, da pogledamo, kaj vse se je zgodilo v tem času v fiziki osnovnih fermionskih in bozonskih polj in v kozmologiji in kaj so novega v tem času ponudile meritve skupaj s predlaganimi teorijami. Tehnologija in računalništvo sta medtem presenetljivo hitro napredovala in omogočila skoraj neverjetne meritve na vseh področjih fizike, tudi ali še posebej v fiziki fermionskih in bozonskih polj ter v kozmologiji. Poskusi so potrdili predpostavke standardnega modela ne da bi prinesli kakršnekoli presenečenje. Zadnje polje, ki ga predpostavi standardni model, Higgsov skalar, ki je bil odkrit junija 2012, so potrdili v marcu 2013. Vedno bolj natančne meritve mas kvarkov in leptonov in antikvarkov in antileptonov, mešalnih matrik kvarkov in leptonov, mase Higgsovega skalarja, vezanih stanj kvarkov in leptonov, ponujajo nove in nove podatke. Valovanje gravitacijskega polja smo zaznali februarja 2016 in spet 2017.

Če pogledamo zbirko odprtih vprašanj, ki sva si jih s Holgerjem zastavila pred začetkom delavnic in ki smo jo ob vsaki delavnici sproti dopolnjevali, kaže, da ves čas iščemo odgovor na bistveno vprašanje: Kaj je naslednji korak, ki bo presegel standardni model in bo ponudil ne le razlago in razumevanje za vse predpostavljene lastnosti za kvarke in leptone in antikvarke in antileptone in za vsa doslej opažena bozonska polja, skupaj s Higgsovim skalarjem, ampak tudi za vse opažene pojave v vesolju, kot je na primer hitrost širjenja vesolja, pojav temne snovi, pomen singularnosti črnih luknenj kot kvantnega skupka fermionov in antifermionov in vseh bozonskih polj, vključno z gravitacijskim poljem v drugi kvantizaciji, kako je z nujnostjo obstoja temne energije in mnogih drugih opaženj.

Ko poskušamo razumeti kvantno naravo fermionskih in bozonskih polj, iščemo teorijo, ki je brez anomalij in taka, da nam da končne prispevke za predlagane meritve.

Če se izkaže, da lahko uresničimo te zahteve le, če dovolimo, da ima prostor-čas več kot le opazljive (3 + 1) razsežnosti, potem moramo odgovoriti na vprašanje, zakaj so vse razsežnosti razen $d = (3 + 1)$ skrite. Kako se tedaj v $d = (3 + 1)$ manifestirajo notranji prostori fermionskih in bozonskih polj? In ali je taka teorija lahko brez anomalij in ponudi končne prispevke za obravane dogodke?

Merjenja hitrosti rotacije zvezd okoli centra galaksije in gibanja galaksij v jatah so ena najmočnejših dokazov, da ni le običajna snov, iz katere so zvezde, tista, ki določa lastnosti galaksij in jat galaksij. Tudi neposredne meritve eksperimenta Dama/Libra z vse večjo natančnostjo dokazujejo, da je izmerjeni letni odvisnosti števila dogodkov potrebno verjeti in da te dogodke povzroča temna snov, kakršenkoli že je njen izvor.

Meritve Hubblove konstante navajajo, kako hitro se vesolje širi. Kozmološka merjenja kažejo, da črne luknje obstajajo.

Naše delavnice so ponudile inovativne predloge:

i. Za razlago predpostavk standardnega modela, večinoma z novimi nepojasnjenimi predpostav-
kami.

ii. Za to, iz česa je temna snov.

iii. Za pojav in vzrok inflacije vesolja.

iv. Za to, kaj določa notranji prostor fermionov in bozonov.
v. Kako se izogniti anomalijam in oblikovati kvantne teorije polj, ki jih je mogoče renormalizirati.
vi. Kako predlagati poskuse, ki bi pokazali kaj je naslednji korak po standardnem modelu.
vii. Kako se "Narava odloči" zlomiti simetrije od začetne v razsežnosti d do opaženih v d=(3+1)?
viii. Zakaj je metrika prostora-časa metrika Minkovskega in kako je izbira metrike povezana z razvojem našega(ih) vesolja(ij)? ... In mnogo drugih.

v objavljenih zbornikih in v pozneje objavljenih člankih v revijah so prispevki, ki razpravljajo o teh problemih in ponudijo predloge, kako probleme rešiti. V zadnjih dveh letih Covida-19 je delavnica preko "ZOOM-a" nadomestila naše običajne delavnice. Treba je priznati, da srečanja preko interneta ne morejo nadomestiti pravih srečanj, kjer so vsa vprašanja dobrodošla, tudi če je za odgovore potrebno mnogo časa.

Pogovori in razprave na naših delavnicah sploh niso običajne diskusije. Razprave trajajo po več ur, razdeljene v dvourne bloke, z veliko vprašanji, razlagami, poskusi strinjanja ali nestrinjanja udeležencev in pojasnjevalca.

Čeprav so naše delavnice po internetu omogočile predstavitve del v več nadaljevanjih, je prisotnost v istem prostoru za prave razprave veliko bolj učinkovita.

Ta zadnja, jubilejna delavnica, je potekala delno na Bledu in delno po "ZOOM-u".

Teme, predstavljene in obravnavane na tej delavnici, se nanašajo na vse zgoraj
omenjene teme, strnjene v vprašanje "kako razumeti Naravo".

Odgovore smo poskušali najti v več korakih:
• Kako najti naslednji korak, ki bo odgovoril na odprta vprašanja obeh modelov?
• Kako poiskati teorijo, ki bo pojasnila vse privzetke standardnega modela kvarkov in leptonov ter antikvarkov in antileptonov, ki nastopajo v družinah in si izmenjujejo elektromagnetna, šibka, barvna in skalarna polja?
• Kako razložiti doslej opažene kozmološke pojave?
• Kako poiskati renormalizabilno teorijo brez anomalij za vse poznane fermione in njihova umeritvena polja?
• Kako poiskati renormalizabilno teorijo brez anomalij tudi za vezana in sipana stanja fermionov in bozonov?
• Kako simetrije prispevajo k vezanim stanjem?
Kako obravnavati vsa osnovna fermionska in bozonska polja na ekvivalenten način?
• Kako obravnavati fermionska in bozonska polja, če je prostor-čas res štirirazsežen?
• Kako se "Narava odloči" za zlom simetrij od začetne simetrije v d-razsežem prostotu-času do opaženih v d=(3+1)?
• Zakaj je metrika prostora-časa metrika Minkovskega in kako je izbira metrike povezana z razvojem našega vesolja?
• Kaj povzroča inflacijo vesolja?
• Kdaj se pojavi inflacija? Po elektrošibkem prehodu?
• Kaj določa barvno skalo? Iz česa je naše vesolje poleg iz barionov (večinoma iz prve družine kvarkov i leptonov)?
• Kako črne luknje prispevajo k temni snovi?
Kakšna je vloga simetrij v naravi?
Kako narediti poskuse in kako predlagati modele, da ne bi predlagani model
preveč vplival na izmerjene podatke?

Večina prispevkov je "nenavadnih" v tem smislu, da poskušajo najti nove rešitve odprtih problemov.

Zbornik je razdeljen na dva dela. V prvi del so vključena vabljena predavanja, ki so prispela do organizatorjev pravočasno in so bila tudi recenzirana.

V drugem delu so zbrani prispevki, za katere avtorji menijo, da diskusije niso prinesle odločitve, do katere mere se avtorji strinjajo, ali nestrinjajo s predstavljenimi trditvami, ali pa avtorji menijo, da imajo različni pristopi mnogo skupnega ter lahko pripeljejo do novih idej ali novega razumevanja ali celo do sodelovanja.

Nekatere od diskusij, ki so se začele med delavnico, se v tem zborniku ne pojavijo. Morda se bodo nadaljevale na naslednji delavnici in bodo zapisane v naslednjem zborniku.

Organizatorji se iskreno zahvaljujejo vsem sodelujočim na delavnici za učinkovite predstavitve del, za živahne razprave in dobro delovno vzdušje, kljub temu, da je večina udeležencev sodelovala preko spleta, ki ga je vodil Maxim Yu. Khlopov.

Bralec najde vse pogovore in kmalu tudi celoten Zbornik na uradni spletni strani delavnice: http://bsm.fmf.uni-lj.si/bled2022bsm/presentations.html, in na forumu Cosmovia.

*Norma Mankoč Borštnik, Holger Bech Nielsen, Maksim Khlopov, Astri Kleppe*
Ljubljana, decembra 2022

# Invited talks

1. New and recent results, and perspectives from DAMA/LIBRA–phase2
R. Bernabei, P. Belli, A. Bussolotti, V. Caracciolo, R. Cerulli, N. Ferrari, A. Leoncini, V. Merlo, F. Montecchia, F. Cappella, A. d'Angelo, A. Incicchitti, A. Mattei, C.J. Dai, X.H. Ma, X.D. Sheng, Z.P. Ye, [arXiv:2209.00882]

2. Neutrino production by photons scattered on dark matter, V. Beylin, [arXiv:2211.09054,v2]

3. Elusive anomalies, L. Bonora, [arXiv:2207.03279]

4. Maximally Precise Tests of the Standard Model: Elimination of Perturbative QCD Renormalization Scale and Scheme Ambiguities, S. J. Brodsky, [arXiv:2112.02453]

5. Predictions of Additional Baryons and Mesons, Paul H. Frampton, [arXiv: 2207.12408]

6. Possibility of Additional Intergalactic and Cosmological Dark Matter, Paul H. Frampton, [arXiv:2209.05349]

7. Near-inflection point inflation and production of dark matter during reheating, A. Ghoshal, G. Lambiase, S. Pal, A. Paul, S. Porey, [arxiv:2211.15061]

8. Quark masses and mixing from a SU(3) gauge family symmetry, A. Hernandez-Galeana, [Workshop proceedings]

9. Evolution and Possible Forms of Primordial Antimatter and Dark Matter celestial objects, Maxim Yu. Khlopov, O.M. Lecian, [arxiv:2211.09579]

10. The Problem of Particle-Antiparticle in Particle Theory, Felix M Lev, [arxiv:2302.10794]

11. Clifford odd and even objects, offering description of internal space of fermion and boson fields, respectively, open new insight into next step beyond the standard model,
N. S. Mankoč Borštnik, [arxiv:2301.04466]

12. A Unified Solution to the Big Problems of the Standard model, R. N. Mohapatra, N. Okada, [arXiv: 2207.10619]

13. Dusty Dark Matter Pearls Developed, H.B. Nielsen, C. D. Froggatt, [arxiv:2303.06061]

14. What gives a "theory of Initial Conditions?", H.B. Nielsen, K. Nagao, [arxiv:2303.04618]

15. Emergent phenomena in QCD: The holographic perspective, Guy F. de Téramond, [arxiv:2212.14028]

16. Planetary relationship as the new signature from the dark Universe, K. Zioutas, V. Anastassopoulos, A Argiriou, G. Cantatore, S. Cetin, A. Gardikiotis, M. Karuza, A. Kryemadhi, M. Maroudas, A. Mastronikolis, K. Ozbozduman, Y. K. Semertzidis, M. Tsagri, I. Tsagris, [Researchgate.net]

# 18. Abstracts of talks presented at the Workshop and in the Cosmovia forum

19. Discussion of cosmological acceleration and dark energy, Felix M Lev, [arxiv:2302.10794]

20. Clifford odd and even objects in even and odd dimensional spaces, N. S. Mankoč Borštnik, [arXiv: 2210.06256]

21. Modules over Clifford algebras as a basis for the theory of second quantization of spinors, V. V. Monakhov, [Workshop proceedings]

22. A new view on cosmology, with non-translational invariant Hamiltonian, H. B. Nielsen, M. Ninomiya, [arXiv:2303.03796]

23. A new Paradigm for the Dark Matter Phenomenon, P. Salucci, [arxiv:2008.04052]

24. On the construction of artificial empty space, Elia Dmitrieff, [Workshop proceedings]

25. **Virtual Institute of Astroparticle physics as the online platform for studies of BSM physics and cosmology, Maxim Yu. Khlopov**

Abstract. The relaxation of pandemia conditions is not complete and the meetings in person are to be still accompanied by online sessions, leading to their hybrid forms. The unique multi-functional complex of Virtual Institute of Astroparticle Physics (VIA) operating on website http://viavca.in2p3.fr/site.html, provides the platform for online virtual meetings.
We review VIA experience in presentation online for the most interesting theoretical and experimental results, participation online in conferences and meetings, various forms of collaborative scientific work as well as programs of education at distance, combining online videoconferences with extensive library of records of previous meetings and Discussions on Forum. Since 2014 VIA online lectures combined with individual work on Forum acquired the form of Open Online Courses. Aimed to individual work with students the Course is not Massive, but the account for the number of visits to VIA site converts VIA in a specific tool for MOOC activity. VIA sessions, being a traditional part of Bled Workshops' program, have converted at XXV Bled Workshop "What comes beyond the Standard models?" into the hybrid format, combining streaming of the presentations in the Plemelj House (Bled, Slovenia) with distant talks, preserving the traditional creative nonformal atmosphere of Bled Workshop meetings. We openly discuss the state of art of VIA platform.

25.1 **Introduction**

Studies in astroparticle physics link astrophysics, cosmology, particle and nuclear physics and involve hundreds of scientific groups linked by regional networks (like ASPERA/ApPEC [1, 2]) and national centers. The exciting progress in these studies will have impact on the knowledge on the structure of microworld and Universe in their fundamental relationship and on the basic, still unknown, physical laws of Nature (see e.g. [3, 4] for review).

The progress of precision cosmology and experimental probes of the new physics at the LHC and in nonaccelerator experiments, as well as the extension of various indirect studies of physics beyond the Standard model involve with necessity their nontrivial links. Virtual Institute of Astroparticle Physics (VIA) [5] was organized with the aim to play the role of an unifying and coordinating platform for such studies.

Starting from the January of 2008 the activity of the Institute took place on its website [6] in a form of regular weekly video conferences with VIA lectures, covering all the theoretical and experimental activities in astroparticle physics and related topics. The library of records of these lectures, talks and their presentations was accomplished by multi-lingual Forum.

Since 2008 there were 220 VIA online lectures, VIA has supported distant presentations of 192 speakers at 32 Conferences and provided transmission of talks at 78 APC Colloquiums. In 2008 VIA complex was effectively used for the first time for participation at distance in XI Bled Workshop [7]. Since then VIA video conferences became a natural part of Bled Workshops' programs, opening the virtual room of discussions to the world-wide audience. Its progress was presented in [8–20].

Here the current state-of-art of VIA complex, integrated in 2009 - 2022 in the structure of APC Laboratory, is presented in order to clarify the way in which discussion of open questions beyond the standard models of both particle physics and cosmology were supported by the platform of VIA facility at the hybrid Memorial XXV Bled Workshop. The relaxation of the conditions of pandemia, making possible offline meetings, is still not complete, preventing many participants to attend these meetings. In this situation VIA videoconferencing became the only possibility to continue in 2022 traditions of open discussions at Bled meetings combining streams of the offline presentations and support of distant talks and involving distant participants in these discussions.

## 25.2 VIA structure and activity

### 25.2.1 The problem of VIA site

The structure of the VIA site was based on Flash and is virtually ruined now in the lack of Flash support. This original structure is illustrated by the Fig. 25.1. The home page, presented on this figure, contained the information on the coming and records of the latest VIA events. The upper line of menu included links to directories (from left to right): with general information on VIA (About VIA); entrance to VIA virtual rooms (Rooms); the library of records and presentations (Previous), which contained records of VIA Lectures (Previous → Lectures), records of online transmissions of Conferences (Previous → Conferences), APC Colloquiums (Previous → APC Colloquiums), APC Seminars (Previous → APC Seminars) and Events (Previous → Events); Calendar of the past and future VIA events (All events) and VIA Forum (Forum). In the upper right angle there were links to Google search engine (Search in site) and to contact information (Contacts). The announcement of the next VIA lecture and VIA online transmission of APC Colloquium occupied the main part of the homepage with the record of the most recent VIA events below. In the announced time of the event (VIA lecture or transmitted APC Colloquium) it was sufficient to click on "to participate" on the announcement and to Enter as Guest (printing your name) in the corresponding Virtual room. The Calendar showed the program of future VIA lectures and events. The right column on the VIA homepage listed the announcements of the regularly up-dated hot news of Astroparticle physics and related areas.

In the lack of Flash support this system of links is ruined, but fortunately, they continue to operate separately and it makes possible to use VIA Forum, by direct link to it, as well as direct inks to virtual room of adobeConnect used for regular Laboratory meetings and

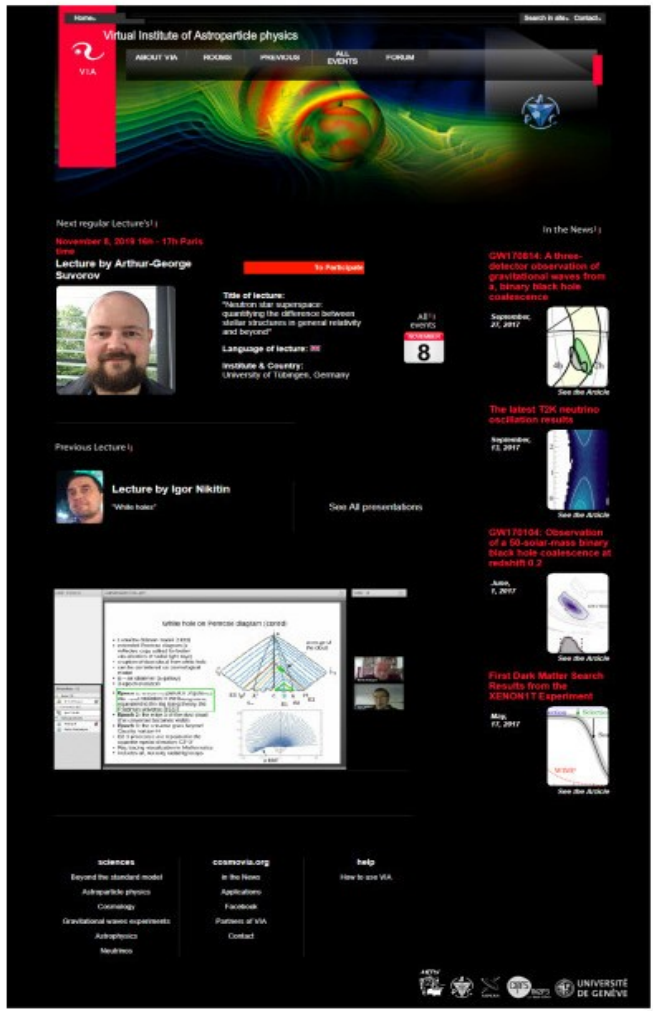

Fig. 25.1: The original home page of VIA site

Seminar and to Zoom (see Fig 25.2). The necessity to restore all the links within VIA complex is a very important task to revive the full scale of VIA activity. Another problem is the necessity to convert .flv files of records in mp4 format.

### 25.2.2 VIA activity

In 2010 special COSMOVIA tours were undertaken in Switzerland (Geneva), Belgium (Brussels, Liege) and Italy (Turin, Pisa, Bari, Lecce) in order to test stability of VIA online transmissions from different parts of Europe. Positive results of these tests have proved the stability of VIA system and stimulated this practice at XIII Bled Workshop. The records of the videoconferences at the XIII Bled Workshop were put on VIA site [21].
Since 2011 VIA facility was used for the tasks of the Paris Center of Cosmological Physics (PCCP), chaired by G. Smoot, for the public program "The two infinities" conveyed by J.L.Robert and for effective support a participation at distance at meetings of the Double Chooz collaboration. In the latter case, the experimentalists, being at shift, took part in the collaboration meeting in such a virtual way.
The simplicity of VIA facility for ordinary users was demonstrated at XIV Bled Workshop in 2011. Videoconferences at this Workshop had no special technical support except for WiFi Internet connection and ordinary laptops with their internal webcams and microphones.
This test has proved the ability to use VIA facility at any place with at least decent Internet connection. Of course the quality of records is not as good in this case as with the use of special equipment, but still it is sufficient to support fruitful scientific discussion as can be illustrated by the record of VIA presentation "New physics and its experimental probes" given by John Ellis from his office in CERN (see the records in [22]).

In 2012 VIA facility, regularly used for programs of VIA lectures and transmission of APC Colloquiums, has extended its applications to support M.Khlopov's talk at distance at Astrophysics seminar in Moscow, videoconference in PCCP, participation at distance in APC-Hamburg-Oxford network meeting as well as to provide online transmissions from the lectures at Science Festival 2012 in University Paris7. VIA communication has effectively resolved the problem of referee's attendance at the defence of PhD thesis by Mariana Vargas in APC. The referees made their reports and participated in discussion in the regime of VIA videoconference. In 2012 VIA facility was first used for online transmissions from the Science Festival in the University Paris 7. This tradition was continued in 2013, when the transmissions of meetings at Journées nationales du Développement Logiciel (JDEV2013) at Ecole Politechnique (Paris) were organized [24].

In 2013 VIA lecture by Prof. Martin Pohl was one of the first places at which the first hand information on the first results of AMS02 experiment was presented [23].

In 2014 the 100th anniversary of one of the foundators of Cosmoparticle physics, Ya. B. Zeldovich, was celebrated. With the use of VIA M.Khlopov could contribute the programme of the "Subatomic particles, Nucleons, Atoms, Universe: Processes and Structure International conference in honor of Ya. B. Zeldovich 100th Anniversary" (Minsk, Belarus) by his talk "Cosmoparticle physics: the Universe as a laboratory of elementary particles" [25] and the programme of "Conference YaB-100, dedicated to 100 Anniversary of Yakov Borisovich Zeldovich" (Moscow, Russia) by his talk "Cosmology and particle physics".

In 2015 VIA facility supported the talk at distance at All Moscow Astrophysical seminar "Cosmoparticle physics of dark matter and structures in the Universe" by Maxim Yu. Khlopov and the work of the Section "Dark matter" of the International Conference on Particle Physics and Astrophysics (Moscow, 5-10 October 2015). Though the conference room was situated in Milan Hotel in Moscow all the presentations at this Section were given at distance (by Rita Bernabei from Rome, Italy; by Juan Jose Gomez-Cadenas, Paterna, University of Valencia, Spain and by Dmitri Semikoz, Martin Bucher and Maxim Khlopov from Paris) and its proceeding was chaired by M.Khlopov from Paris. In the end of 2015 M. Khlopov gave his distant talk "Dark atoms of dark matter" at the Conference "Progress of Russian Astronomy in 2015", held in Sternberg Astronomical Institute of Moscow State University.

In 2016 distant online talks at St. Petersburg Workshop "Dark Ages and White Nights (Spectroscopy of the CMB)" by Khatri Rishi (TIFR, India) "The information hidden in the CMB spectral distortions in Planck data and beyond", E. Kholupenko (Ioffe Institute, Russia) "On recombination dynamics of hydrogen and helium", Jens Chluba (Jodrell Bank Centre for Astrophysics, UK) "Primordial recombination lines of hydrogen and helium", M. Yu. Khlopov (APC and MEPHI, France and Russia)"Nonstandard cosmological scenarios" and P. de Bernardis (La Sapiensa University, Italy) "Balloon techniques for CMB spectrum research" were given with the use of VIA system. At the defense of PhD thesis by F. Gregis VIA facility made possible for his referee in California not only to attend at distance at the presentation of the thesis but also to take part in its successive jury evaluation.

Since 2018 VIA facility is used for collaborative work on studies of various forms of dark matter in the framework of the project of Russian Science Foundation based on Southern Federal University (Rostov on Don). In September 2018 VIA supported online transmission of 17 presentations at the Commemoration day for Patrick Fleury, held in APC.

The discussion of questions that were put forward in the interactive VIA events is continued and extended on VIA Forum. Presently activated in English,French and Russian with trivial extension to other languages, the Forum represents a first step on the way to multi-lingual character of VIA complex and its activity. Discussions in English on Forum are arranged along the following directions: beyond the standard model, astroparticle physics, cosmology, gravitational wave experiments, astrophysics, neutrinos. After each VIA lecture its pdf presentation together with link to its record and information on the discussion during it are put in the corresponding post, which offers a platform to continue discussion in replies to this post.

Fig. 25.2: The current home page of VIA site

### 25.2.3 **VIA e-learning, OOC and MOOC**

One of the interesting forms of VIA activity is the educational work at distance. For the last eleven years M.Khlopov's course "Introduction to cosmoparticle physics" is given in the form of VIA videoconferences and the records of these lectures and their ppt presentations are put in the corresponding directory of the Forum [26]. Having attended the VIA course of lectures in order to be admitted to exam students should put on Forum a post with their small thesis. In this thesis students are proposed to chose some BSM model and to study the cosmological scenario based on this chosen model. The list of possible topics for such thesis is proposed to students, but they are also invited to chose themselves any topic of their own on possible links between cosmology and particle physics. Professor's comments and proposed corrections are put in a Post reply so that students should continuously present on Forum improved versions of work until it is accepted as admission for student to pass exam. The record of videoconference with the oral exam is also put in the corresponding directory of Forum. Such procedure provides completely transparent way of evaluation of students' knowledge at distance.
In 2018 the test has started for possible application of VIA facility to remote supervision of student's scientific practice. The formulation of task and discussion of progress on work are recorded and put in the corresponding directory on Forum together with the versions of student's report on the work progress.Since 2014 the second semester of the course on Cosmoparticle physics is given in English
and converted in an Open Online Course. It was aimed to develop VIA system as a possible accomplishment for Massive Online Open Courses (MOOC) activity [27]. In 2016 not only students from Moscow, but also from France and Sri Lanka attended this course. In 2017 students from Moscow were accompanied by participants from France, Italy, Sri Lanka and India [28]. The students pretending to evaluation of their knowledge must write their small thesis, present it and, being admitted to exam, pass it in English. The restricted number of online connections to videoconferences with VIA lectures is compensated by the wide-world access to their records on VIA Forum and in the context of MOOC VIA Forum and videoconferencing system can be used for individual online work with advanced participants. Indeed Google Analytics shows that since 2008 VIA site was visited by more than 250 thousand visitors from 155 countries, covering all the continents by its geography (Fig. 25.3). According to this statistics more than half of these visitors continued to enter VIA site after the first visit. Still the form of individual educational work makes VIA facility most appropriate for PhD courses and it could be involved in the International PhD program on Fundamental Physics, which was planned to be started on the basis of Russian-French collaborative agreement. In 2017 the test for the ability of VIA to support fully distant education and evaluation of students (as well as for work on PhD thesis and its distant defense) was undertaken. Steve Branchu from France, who attended the Open Online
Course and presented on Forum his small thesis has passed exam at distance. The whole procedure, starting from a stochastic choice of number of examination ticket, answers to ticket questions, discussion by professors in the absence of student and announcement of result of exam to him was recorded and put on VIA Forum [29].
In 2019 in addition to individual supervisory work with students the regular scientific and creative VIA seminar is in operation aimed to discuss the progress and strategy of students scientific work in the field of cosmoparticle physics.
In 2020 the regular course now for M2 students continued, but the problems of adobe Connect, related with the lack of its support for Flash in 2021 made necessary to use the platform of Zoom, This platform is rather easy to use and provides records, as well as whiteboard tools for discussions online can be solved by accomplishments of laptops bygraphic tabloids. In 2022 the Open Online Course for M2 students was accompanied by special course "Cosmoparticle physics", given in English for English speaking M1 students.

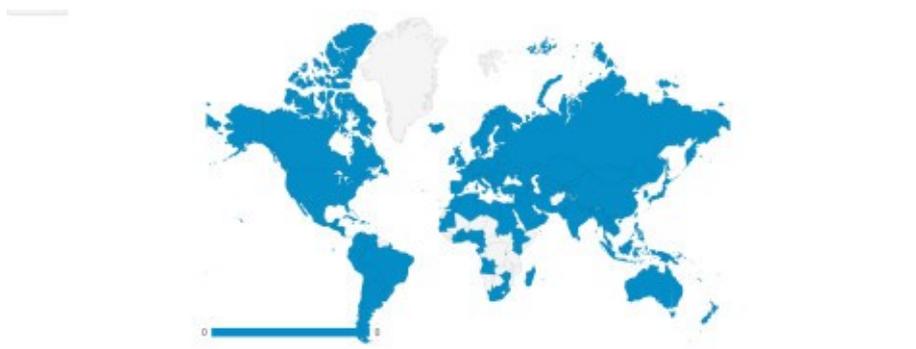

Fig. 25.3: Geography of VIA site visits according to Google Analytics

### 25.2.4 Organisation of VIA events and meetings

First tests of VIA system, described in [5, 7–9], involved various systems of videoconferencing. They included skype, VRVS, EVO, WEBEX, marratech and adobe Connect. In the result of these tests the adobe Connect system was chosen and properly acquired. Its advantages were: relatively easy use for participants, a possibility to make presentation in a video contact between presenter and audience, a possibility to make high quality records, to use a whiteboard tools for discussions, the option to open desktop and to work online with texts in any format. The lack of support for Flash, on which VIA site was originally based, made necessary to use Zoom, which shares all the above mentioned advantages.
Regular activity of VIA as a part of APC included online transmissions of all the APC Colloquiums and of some topical APC Seminars, which may be of interest for a wide audience. Online transmissions were arranged in the manner, most convenient for presenters, prepared to give their talk in the conference room in a normal way, projecting slides from their laptop on the screen. Having uploaded in advance these slides in the VIA system, VIA operator, sitting in the conference room, changed them following presenter, directing simultaneously webcam on the presenter and the audience. If the advanced uploading was not possible, VIA streaming was used - external webcam and microphone are directed to presenter and screen and support online streaming. This experience has found proper place in the current weakening of the pandemic conditions and regular meetings in real can be streamed. Moreover, such streaming can be made without involvement of VIA operator, by direction of webcam towards the conference screen and speaker.

### 25.2.5 VIA activity in the conditions of pandemic

The lack of usual offline connections and meetings in the conditions of pandemia made the use of VIA facility especially timely and important. This facility supports regular weekly meetings of the Laboratory of cosmoparticle studies of the structure and dynamics of Galaxy in Institute of Physics of Southern Federal University (Rostov on Don, Russia) and M.Khlopov's scientific - creative seminar and their announcements occupied their permanent position on VIA homepage (Fig. 25.2), while their records were put in respective place of VIA forum, like [31] for Laboratory meetings.
The platform of VIA facility was used for regular Khlopov's course "Introduction to Cos-moparticle physics" for M2 students of MEPHI (in Russian) and in 2020 supported regular seminars of Theory group of APC.

The programme of VIA lectures continued to present hot news of astroparticle physics and cosmology, like talk by Zhen Cao from China on the progress of LHAASO experiment or lecture by Sunny Vagnozzi from UK on the problem of consistency of different measurements of the Hubble constant.

The results of this activity inspired the decision to hold in 2020 XXIII Bled Workshop online on the platform of VIA [19].

The conditions of pandemia continued in 2021 and VIA facility was successfully used to provide the platform for various online meetings. 2021 was announced by UNESCO as A.D.Sakharov year in the occasion of his 100th anniversary VIA offered its platform for various events commemorating A.D.Sakharov's legacy in cosmoparticle physics. In the framework of 1 Electronic Conference on Universe ECU2021), organized by the MDPI journal "Universe" VIA provided the platform for online satellite Workshop "Developing A.D.Sakharov legacy in cosmoparticle physics" [32].

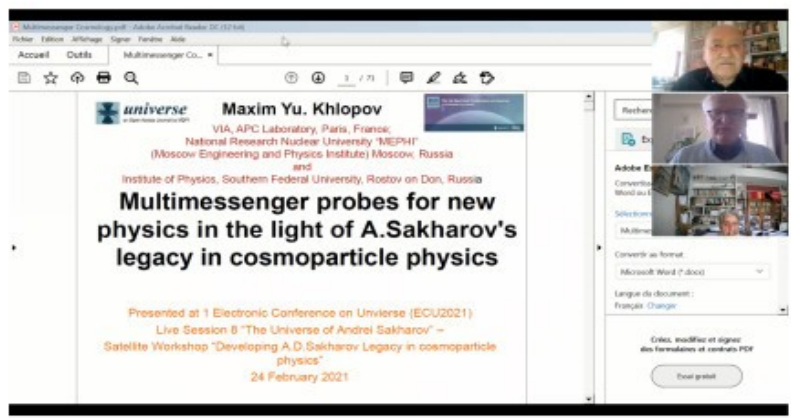

Fig. 25.4: M.Khlopov's talk "Multimessenger probes for new physics in the light of A.D.Sakharov legacy in cosmoparticle physics" at the satellite Workshop "Developing A.D.Sakharov legacy in cosmoparticle physics" of ECU2021.

### 25.3 VIA platform at Hybrid Memorial XXV Bled Workshop

VIA sessions at Bled Workshops continued the tradition coming back to the first experience at XI Bled Workshop [7] and developed at XII, XIII, XIV, XV, XVI, XVII, XVIII, XIX, XX, XXI and XXII Bled Workshops [8–18]. They became a regular but supplementary part of the Bled Workshop's program. In the conditions of pandemia it became the only form of Workshop activity in 2020 [19] and in 2021 [20].

During the XXV Bled Workshop the announcement of VIA sessions was put on VIA home page, giving an open access to the videoconferences at the Workshop sessions. The preliminary program as well as the corrected program for each day were continuously put on Forum with the slides and records of all the talks and discussions [33].

VIA facility tried to preserve the creative atmosphere of Bled discussions. The program of XXV Bled Workshop combined talks presented in Plemelj House in Bled, which were streamed by VIA facility, as the talk "Understanding nature with the spin-charge- family theory, New way of second quantization of fermions and bosons" by Norma Mankoc-Borstnik (Fig. 25.5) with talks given in the format videoconferences "Recent results and empowered perspectives of DAMA/LIBRA-phase2" by R.Bernabei, (Fig. 25.6), from Rome University, Italy (see records in [33]).

During the Workshop the VIA virtual room was open, inviting distant participants to join the discussion and extending the creative atmosphere of these discussions to the worldwide audience. The participants joined these discussions from different parts of world. The talk "A Nietzschean paradigm for the dark matter phenomenon" was given by Paolo Salucci from Italy (Fig. 25.7). R.Mohapatra and S. Brodsky gave their talks from US and E.Kiritsis from from Paris. The online talks were combined with presentations in Bled such as "Dusty dark matter pearls developed" by H.B. Nielsen (Fig. 25.8) or "Cosmological reflection of the BSM physics" by M.Y. Khlopov (Fig. 25.9).

The distant VIA talks highly enriched the Workshop program and streaming of talks from Bled involved distant participants in fruitful discussions. The use of VIA facility has provided remote presentation of students' scientific debuts in BSM physics and cosmology. The records of all the talks and discussions can be found on VIA Forum [33].
VIA facility has managed to join scientists from Mexico, USA, France, Italy, Russia, Slovenia, India and many other countries in discussion of open problems of physics and cosmology beyond the Standard models.

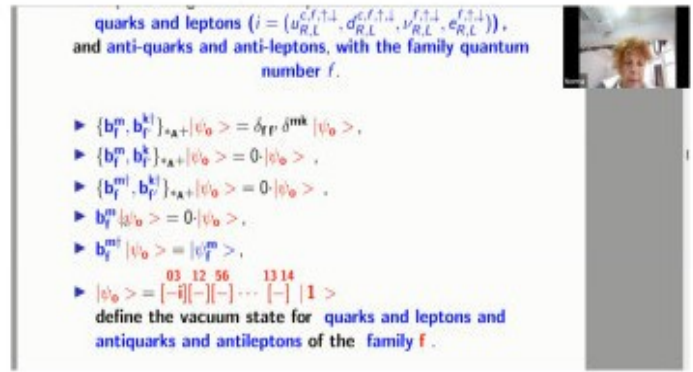

Fig. 25.5: VIA stream of the talk "Understanding nature with the spin-charge-family theory, New way of second quantization of fermions and bosons" by Norma Mankoc-Borstnik at XXV Bled Workshop

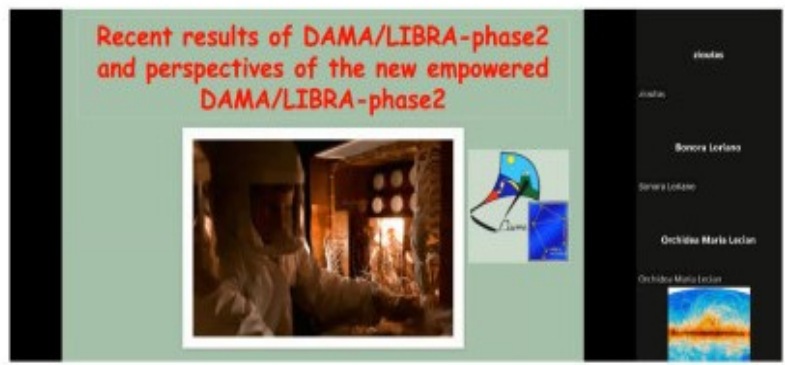

Fig. 25.6: VIA talk "Recent results and empowered perspectives of DAMA/LIBRA-phase2" by R.Bernabei from Rome at XXV Bled Workshop

25.4 **Conclusions**

The Scientific-Educational complex of Virtual Institute of Astroparticle physics provides regular communication between different groups and scientists, working in different scientific fields and parts of the world, the first-hand information on the newest scientific results, as well as support for various educational programs at distance. This activity would easily allow finding mutual interest and organizing task forces for different scientific topics of cosmology, particle physics, astroparticle physics and related topics. It can help in the elaboration of strategy of experimental particle, nuclear, astrophysical and cosmological studies as well as in proper analysis of experimental data. It can provide young talented people from all over the world to get the highest level education, come in direct interactive contact with the world known scientists and to find their place in the fundamental research.
These educational aspects of VIA activity can evolve in a specific tool for International PhD program for Fundamental physics. Involvement of young scientists in creative discussions
was an important aspect of VIA activity at XXV Bled Workshop. VIA applications can go far beyond the particular tasks of astroparticle physics and give rise to an interactive system of mass media communications.

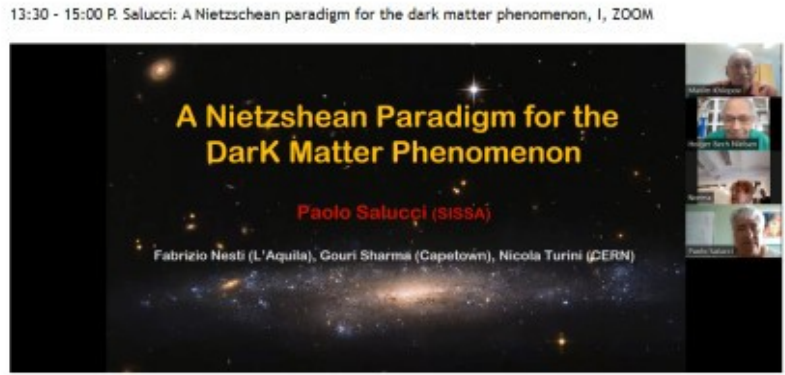

Fig. 25.7: VIA talk "A Nietzschean paradigm for the dark matter phenomenon" by Paolo Salucci at XXV Bled Workshop

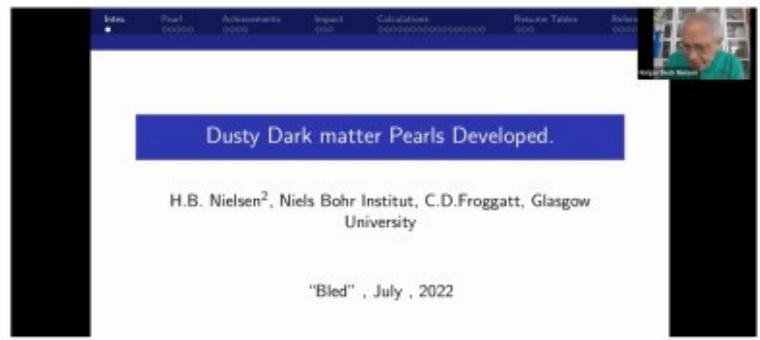

Fig. 25.8: VIA stream of talk "Dusty dark matter pearls developed" by Holger Bech Nielsen at XXV Bled Workshop

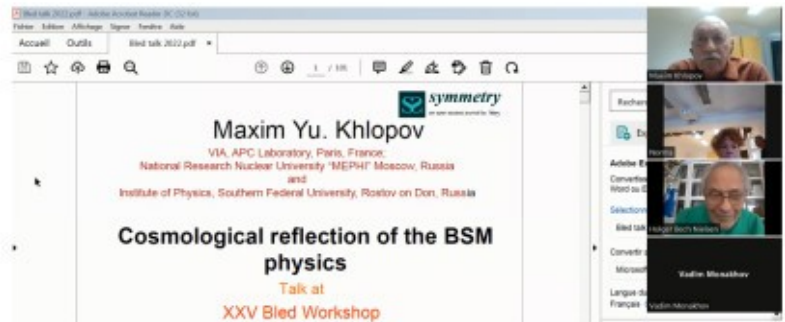

Fig. 25.9: VIA stream of talk "Cosmological reflection of the BSM physics" by Maxim Yu. Khlopov at XXV Bled Workshop

VIA sessions, which became a natural part of a program of Bled Workshops, maintained in 2022 the platform for online discussions of physics beyond the Standard Model involving distant participants from all the world in the fruitful atmosphere of Bled offline meeting. This discussion can continue in posts and post replies on VIA Forum. The experience of VIA applications at Bled Workshops plays important role in the development of VIA facility as an effective tool of e-science and e-learning.

One can summarize the advantages and flaws of online format of Bled Workshop. It makes possible to involve in the discussions scientists from all the world (young scientists, especially) free of the expenses related with meetings in real (voyage, accommodation, ...), but loses the advantage of nonformal discussions at walks along the beautiful surrounding of the Bled lake and other places of interest. The improvement of VIA technical support by involvement of Zoom provided better platform for nonformal online discussions, but in no case can be the substitute for offline Bled meetings and its creative atmosphere in real, which has revived at the offline XXV Bled Workshop. One can summarize that VIA facility provides important tool of the offline Bled Workshop, involving world-wide participants in its creative and open discussions.

**Acknowledgements**

The initial step of creation of VIA was supported by ASPERA. I express my tribute to memory of P.Binetruy and S.Katsanevas and express my gratitude to J.Ellis for permanent stimulating support, to J.C. Hamilton for early support in VIA integration in the structure of APC laboratory, to K.Belotsky, A.Kirillov, M.Laletin and K.Shibaev for assistance in educational VIA program, to A.Mayorov, A.Romaniouk and E.Soldatov for fruitful collaboration, to K.Ganga, J.Errard, A.Kouchner and D.Semikoz for collaboration in development of VIA activity in APC, to M.Pohl, C. Kouvaris, J.-R.Cudell, C. Giunti, G. Cella, G. Fogli and F. DePaolis for cooperation in the tests of VIA online transmissions in Switzerland, Belgium and Italy and to D.Rouable for help in technical realization and support of VIA complex. The work was financially supported by Southern Federal University, 2020 Project VnGr/2020-03-IF. I express my gratitude to the Organizers of Bled Workshop N.S. Manko ̌c Bor ̌stnik, A.Kleppe and H.Nielsen or cooperation in the organization of VIA online Sessions at XXV Bled Workshop. I am grateful to T.E.Bikbaev for technical assistance and help. I am grateful to Sandi Ogrizek for creation of compact links to VIA Forum.

References

1. http://www.aspera-eu.org/
2. http://www.appec.org/
3. M.Yu. Khlopov: Cosmoparticle physics, World Scientific, New York -London-Hong Kong - Singapore, 1999.
4. M.Yu. Khlopov: Fundamentals of Cosmic Particle Physics, CISP-Springer, Cambridge, 2012.
5. M. Y. Khlopov, Project of Virtual Institute of Astroparticle Physics, arXiv:0801.0376 [astro-ph].
6. http://viavca.in2p3.fr/site.html
7. M. Y. Khlopov, Scientific-educational complex - virtual institute of astroparticle physics, 981–862008.
8. M. Y. Khlopov, Virtual Institute of Astroparticle Physics at Bled Workshop, 10177–1812009.
9. M. Y. Khlopov, VIA Presentation, 11225–2322010.
10. M. Y. Khlopov, VIA Discussions at XIV Bled Workshop, 12233–2392011.
11. M. Y. .Khlopov, Virtual Institute of astroparticle physics: Science and education online, 13183–1892012.
12. M. Y. .Khlopov, Virtual Institute of Astroparticle physics in online discussion of physics beyond the Standard model, 14223–2312013.
13. M. Y. .Khlopov, Virtual Institute of Astroparticle physics and "What comes beyond the Standard model?" in Bled, 15285-2932014.
14. M. Y. .Khlopov, Virtual Institute of Astroparticle physics and discussions at XVIII Bled Workshop, 16177-1882015.
15. M. Y. .Khlopov, Virtual Institute of Astroparticle Physics — Scientific-Educational Platform for Physics Beyond the Standard Model, 17221-2312016.
16. M. Y. .Khlopov: Scientific-Educational Platform of Virtual Institute of Astroparticle Physics and Studies of Physics Beyond the Standard Model, 18273-2832017.
17. M. Y. .Khlopov: The platform of Virtual Institute of Astroparticle physics in studies of physics beyond the Standard model, 19383-3942018.
18. M. Y. .Khlopov: The Platform of Virtual Institute of Astroparticle Physics for Studies of BSM Physics and Cosmology, Journal20249-2612019.
19. M. Y. .Khlopov: Virtual Institute of Astroparticle Physics as the Online Platform for Studies of BSM Physics and Cosmology, Journal21249-2632020.
20. M. Y. .Khlopov: Challenging BSM physics and cosmology on the online platform of Virtual Institute of Astroparticle physics, Journal22160-1752021.
21. http : //viavca.in2p3.fr/whatcomesbeyondthestandardmodelsxiii.html
22. http : //viavca.in2p3.fr/whatcomesbeyondthestandardmodelsxiv.html
23. http : //viavca.in2p3.fr/pohlmartin.html
24. In http://viavca.in2p3.fr/ Previous - Events - JDEV 2013
25. http : //viavca.in2p3.fr/zeldovich100meeting.html
26. In https://bit.ly/bled2022bsm Forum- Discussion in Russian - Courses on Cosmoparticle physics
27. In https://bit.ly/bled2022bsm Forum - Education - From VIA to MOOC
28. In https://bit.ly/bled2022bsm Forum - Education - Lectures of Open Online VIA Course 2017
29. In https://bit.ly/bled2022bsm Forum - Education - Small thesis and exam of Steve

Branchu
30. http : //viavca.in2p3.fr/johnellis.html
31. In https://bit.ly/bled2022bsm Forum - LABORATORY OF COSMOPARTICLE STUDIES OF STRUCTURE AND EVOLUTION OF GALAXY
32. In https://bit.ly/bled2022bsm Forum - CONFERENCES - CONFERENCES ASTROPARTICLE PHYSICS - The Universe of A.D. Sakharov at ECU2021
33. In https://bit.ly/bled2022bsm Forum - CONFERENCES BEYOND THE STANDARD MODEL - XXV Bled Workshop "What comes beyond the Standard Model"

# A poem

When I met Elia Dmitrieff in Bled, he told me that he has a datcha by the Lake Baikal.
"Really?" I said, "o, I would love to visit you and see the Baikal!"
He smiled and nodded, and I thought of the Trans-Siberian railway, and made a vague plan
to one day visit him. Some years later, when the war in Ukraine started, I was in total shock, and
soon thereafter, I went into a depression. I thought of the Ukrainians, and I thought of Elia, and I
thought of all my Russian friends. And then I wrote this poem.

*Voyage to Irkutsk*

I thought I one day
would pay you a visit, take the train
many days
through the large waste land woods

We would go to your datcha
chop some wood, make a fire,
drink our tea
and discuss the meaning of charge!

By the hearth you would tell me
your binary codes
of the innermost parts, of your
Glasperlenspiel,

while over the house the moon
would sail,
and mirror its light
in the Lake Baikal,
And you'd say: If you pray, pray
for healing and reason, and if
you believe,
believe
in the good!

I will, I would say, but I want you to tell me
a story where all goes well!
I want it to be in a grandiose land
where oceans are pure, since pollution is banned.

And there must be a house with a garden and trees
with apples and plums, and flowers and bees
And children shall play in the neighbouring wood
and nobody barks, and people are good!

Do you hear, I will come
and visit you soon! Then we'll talk
about binary codes.
When the devils
are gone and the light is restored,
and the roads are repaired
and the flowers all bloom,
I will visit you soon, very soon!

*Astri Kleppe*